\documentclass[12pt]{article}

\usepackage{amsfonts,latexsym,eucal,color,graphicx,epsfig,amssymb,amsmath,booktabs,multirow}

\newcommand{\be}{\begin{eqnarray}}
\newcommand{\ee}{\end{eqnarray}}
\newcommand{\beq}{\begin{eqnarray}}
\newcommand{\eeq}{\end{eqnarray}}
\newcommand{\pd}{\partial}
\newcommand{\nb}{\nabla}
\newcommand{\nn}{\nonumber}

\newcommand{\dalm}{\kern1pt\vbox{\hrule height 0.9pt\hbox{\vrule width 0.9pt\hskip 2.5pt\vbox{\vskip 5.5pt}\hskip 3pt\vrule width 0.3pt}\hrule height 0.3pt}\kern1pt}
\newcommand{\ie}{{\it i.e.}}
\newcommand{\eg}{{\it e.g.}}

\newcommand{\et}{{\it et al.}}

\makeatletter

\@addtoreset{equation}{section}
\makeatletter
\makeatletter
\usepackage{mathrsfs}

 \def\com#1{}
\def\intro#1{} \def\torikeshi#1{}
\begin{document}

\begin{titlepage}

\begin{flushright}
{\small {\tt arXiv:1108.0210 [gr-qc]}} \\
\end{flushright}
\vspace{2cm}
\begin{center}
{\Large {\bf Naked Singularity Explosion in Higher Dimensions}} \\
\vspace{1.5cm}
{\bf Umpei Miyamoto,${}^{1,\ast}$\quad Hiroya Nemoto,${}^1$\quad Masahiro Shimano${}^{1,2}$}\\
\vspace{.5cm} 
{\small {\it ${}^1$Department of Physics, Rikkyo University, Tokyo 171-8501, Japan}}\\
{\small {\it ${}^2$Jumonji Junior and Senior High School, Toshima, Tokyo 170-0004, Japan}}\\
\vspace{.5cm}
{\small {\tt ${}^\ast$umpei@rikkyo.ac.jp}}\\
\vspace{1.2cm}
\end{center}

\begin{abstract}
Motivated by the recent argument that in the TeV-scale gravity trans-Planckian domains of spacetime as effective naked singularities would be generated by high-energy particle (and black-hole) collisions, we investigate the quantum particle creation by naked-singularity formation in general dimensions. Background spacetime is simply modeled by the self-similar Vaidya solution, describing the spherical collapse of a null dust fluid. In a generic case the emission power is found to be proportional to the quadratic inverse of the remaining time to a Cauchy horizon, as known in four dimensions. On the other hand, the power is proportional to the quartic inverse for a critical case in which the Cauchy horizon is `degenerate'. According to these results, we argue that the backreaction of the particle creation to gravity will be important in particle collisions, in contrast to the gravitational collapse of massive stellar objects, since the bulk of energy is carried away by the quantum radiation even if a quantum gravitational effect cutoff the radiation just before the appearance of naked singularity.
\end{abstract}

\end{titlepage}

\tableofcontents

%----------------------------------------------------------------------%
%----------------------------------------------------------------------%
\section{Introduction}
\label{sec:into}
%----------------------------------------------------------------------%
%----------------------------------------------------------------------%

The higher-dimensional scenarios with large~\cite{ArkaniHamed:1998rs} and warped~\cite{Randall:1999ee} extra dimensions were proposed to resolve or reformulate the hierarchy between the gravitational and electroweak interactions. In these scenarios the $d$-dimensional ($d>4$) Planck mass $M_P$ is related to that in 4-dimension $M_{P(4)} (\sim 10^{19} \; {\rm GeV})$ by $ M_{P(4)}^2 \sim L^{d-4} M_P^{d-2} $, where $L$ is the size of an extra dimension. Thus, if $L$ is large enough, the fundamental Planck mass $M_P$ can be as low as a few TeV. For instance, such a small Planck mass is realized if $L \sim 0.1$ cm when $d=6$, $L \sim 10^{-7}$ cm when $d=7$. If the Standard Model particles except the gravitons (and possibly other unobserved particles) are confined to our 3-brane these scenarios are consistent with all current observations.

One of the most striking predictions of such scenarios is the productions of a large number of mini black holes in high-energy particle collisions~\cite{Argyres:1998qn}. A simplified picture of the black-hole production can be put in the following way. The size of a black hole is characterized by Schwarzschild event-horizon radius $r_{\rm eh}$, scaling with its mass as $r_{\rm eh} \sim M^{1/(d-3)}$. If colliding particles have a center-of-mass energy $ M$ above a threshold energy of the order of $M_P$ and an impact parameter less than the Schwarzschild radius, a black hole of mass $M$ is produced. In other words, the total cross section of black-hole production is given by $\sigma_{BH} \simeq \pi r_{\rm eh}^2$. The black holes so produced will decay thermally via the Hawking radiation~\cite{Hawking:1974sw} and be detected in terrestrial collider experiments such as CERN Large Hadron Collider and in ultrahigh-energy cosmic rays. Such possibilities have been extensively studied and known to give rise to rich phenomenology (\eg, see \cite{Kanti:2008eq} for a review).

Recently, one of the present authors and his collaborators pointed out another possibility~\cite{Nakao:2010az} in TeV-gravity scenarios. They argued that effective naked singularities called {\it the visible borders of spacetime} would be generated by high-energy particle collisions. A border of spacetime, originally proposed in \cite{Harada:2004mv}, is defined as a domain of spacetime where the curvature becomes trans-Planckian and acts as an border (or boundary) of classical sub-Planckian domains. A simplified picture of the generation of a visible border can be put in the following way. Suppose colliding particles have a center-of-mass energy above $M_P$, just like the black-hole production mentioned above. Then, suppose that the impact parameter is small enough to make the energy density of the colliding region become trans-Planckian but the impact parameter is {\it larger} than the Schwarzschild radius. If such a situation is possible, the curvature around the colliding region becomes trans-Planckian through the Einstein equation, but the horizon will not form. Therefore, the trans-Planckian domain of spacetime, \ie, the border of spacetime, is visible or naked to outer observers. Paper~\cite{Nakao:2010az} showed by a simple dimensional argument that such phenomena can occur in collider experiments, which is regarded as an effective violation of {\it the cosmic censorship hypothesis}~\cite{Penrose:1969pc} in higher dimensions. In such a visible-border production, in contrast to the black-hole production, the trans-Planckian domain is exposed to observers, and therefore an arena of quantum gravity could be provided. Furthermore, quite recently Okawa, Nakao, and Shibata~\cite{Okawa:2011fv} showed by a fully general relativistic simulation that trans-Planckian domains of spacetime not covered by horizons are produced in the course of black-hole collision in 5 dimensions, strongly supporting the argument in~\cite{Nakao:2010az}.

It would be fare to say that we do not have any rigorous quantum theory of gravity to predict phenomena near the Planckian regime. Nevertheless, it would be important to predict possible phenomena with classical and semiclassical tools available. The study of semiclassical effects during naked-singularity formation (in 4 dimensions) has a relatively long history, which dates back to the seminal works by Ford and Parker~\cite{Ford:1978ip} and by Hiscock, Williams, and Eardley~\cite{Hiscock:1982pa}. (An incomplete list of studies in this direction is~\cite{Barve:1998ad,Singh:2000sp,Harada:2000me,Miyamoto:2003wr,Miyamoto:2004ba}. See \cite{Harada:2001nj} for a review.) Typically, when a (strong) globally naked singularity forms, which violates the weak version of the cosmic censorship hypothesis~\cite{Penrose:1969pc}, the power of particle creation diverges at the Cauchy horizon (if one neglects the backreactions to spacetime).

Due to a universality of the particle creation by naked singularities, one can expect that similar phenomena occur in higher dimensions, which were addressed by the present authors for the first time in~\cite{Miyamoto:2010vn} and are being analyzed further in this paper. We should note several points prior to modeling. Firstly, the generation of visible borders or naked singularities in large-extra-dimension scenarios would be a highly asymmetric phenomenon. Namely, the gravity propagates in every direction; the Standard Model particles are confined to our 3-brane; the non-zero impact parameters of particle collisions are essential. It seems not so easy to model the visible-border formation by known exact solutions nor numerical solutions to the Einstein equation. Therefore, in this paper as well as in the previous paper~\cite{Miyamoto:2010vn}, we adopt the Vaidya solution as a first step, describing the spherically symmetric naked-singularity formation due to the accretion of a null dust fluid. Secondly, it is certain that the spectra of particle creation will provide important informations in order to identity what the products of collision are in experiments. However, there is a fundamental problem in estimating the spectrum of created particles (\ie, the Bogoliubov coefficients) when the singularity is globally naked: we do not know how to impose boundary conditions on a quantum field at the singularity. Thus, in this paper we focus only on the power and energy emitted, which can be evaluated locally at the price of having no information of the spectrum.\footnote{In the previous work~\cite{Miyamoto:2010vn}, in order to avoid such an ambiguity of boundary conditions and calculate the spectrum we adopted the model describing the formation of a {\it marginally} naked singularity, in which the singularity is observable only within the event horizon (so, it violates only the strong version of the cosmic censorship hypothesis). In such a spacetime the Cauchy horizon and event horizon coincide, and therefore one has to impose no boundary conditions at the singularity. However, there is no physical reason {\it a priori} that such a particular causal structure is preferred. Thus, in this paper we investigate the particle creation in globally naked singularities, which is more generic than the marginally naked singularity.}

The organization of this paper is as follow. In the next section, we introduce the self-similar Vaidya solution. Then, in section~\ref{sec:null} we obtain the null geodesics in the Vaidya spacetime, which is essential to estimate the particle creation with the geometric-optics approximation. In section~\ref{sec:power} we evaluate the power and energy of particle creation. Section~\ref{sec:conc} is devoted to discussions. Some calculations are relegated to appendices. We work in the Planck units, in which $c=G=\hbar=1$ ($G$ is the $d$-dimensional gravitational constant), otherwise noted.

%----------------------------------------------------------------------%
%----------------------------------------------------------------------%
\section{Self-similar Vaidya collapse}
\label{sec:BG}
%----------------------------------------------------------------------%
%----------------------------------------------------------------------%

We consider the $d$-dimensional ($ d \geq 4 $) spherically symmetric collapse of a dust fluid whose line element is given by
\be
	ds^2
	=
	-\left(
		1-\frac{2m(v)}{r^{d-3}}
	\right) dv^2 + 2 dv dr + r^2 d\Omega_{d-2}^2,
\label{metric}
\ee
where $d\Omega_{d-2}^2$ is the line element of a unit ($d-2$)-sphere. We assume the following form of the mass function
\be
	m(v)
	=
	\left\{
	\begin{array}{llc}
		0, & v < 0 & (\mbox{region I})\\
		\mu v^{d-3}, & 0 \leq v < v_0 & (\mbox{region II})\\
		\mu v_0^{d-3}, & v \geq v_0 & (\mbox{region III})
	\end{array}
	\right. .
\label{m}
\ee
Namely, the dust fluid begins to infall toward the center at $v=0$. Constant $\mu$  ($>0$) represents an accretion rate. The above specific form of $m(v)$ in the region II assures that the spacetime is self-similar or homothetic. Then, the infalling stops at $v=v_0$ and the outer spacetime is described by the Schwarzschild-Tangherlini solution with the mass parameter $M := \mu v_0^{d-3}$.\footnote{A physical mass (ADM mass) defined in the asymptotic region is given by 
$ M_{\rm phys} =  (d-2)\Omega_{d-2} M / 8\pi $, where $\Omega_{d-2}=2\pi^{(d-1)/2}/\Gamma[(d-1)/2]$ is the volume of unit ($d-2$)-sphere~\cite{Myers:1986un}.
} The Kretschmann invariant is calculated as
\be
	R^{\alpha\beta\gamma\delta} R_{\alpha\beta\gamma\delta}
	=
	4(d-1)(d-2)^2 (d-3) \frac{ m^2(v) }{ r^{2(d-1)} }.
\ee
Thus, the center ($r=0$) is a curvature singularity unless $m(v)$ vanishes there.

In the region I the spacetime is flat and a usual retarded time is introduced by
\be
	\bar{u} := v - 2r,
\ee
with which the line element is written as
\be
	ds_{\rm I}^2
	=
	-d\bar{u}dv + r^2 d\Omega_{d-2}^2.
\ee
In the region III a retarded time is introduced with the tortoise coordinate $r_\ast$ by
\be
	u := v - 2 r_\ast (r),
\;\;\;
	r_\ast(r) := \int^r \frac{dr}{f(r)},
\;\;\;
	f(r) := 1 - \frac{2M}{r^{d-3}},
\label{U}
\ee
with which the line element is written as
\be
	ds_{\rm III}^2
	=
	-f(r) du dv + r^2 d\Omega_{d-2}^2.
\ee
It is noted that the event horizon is given by $r= r_{\rm eh} := (2\mu)^{1/(d-3)} v_0$ in the region III .

\begin{figure}[bth]
	\begin{center}
		\setlength{\tabcolsep}{ 20 pt }
		\begin{tabular}{ cc }
			(a) & (b) \\
			\includegraphics[width=7cm]{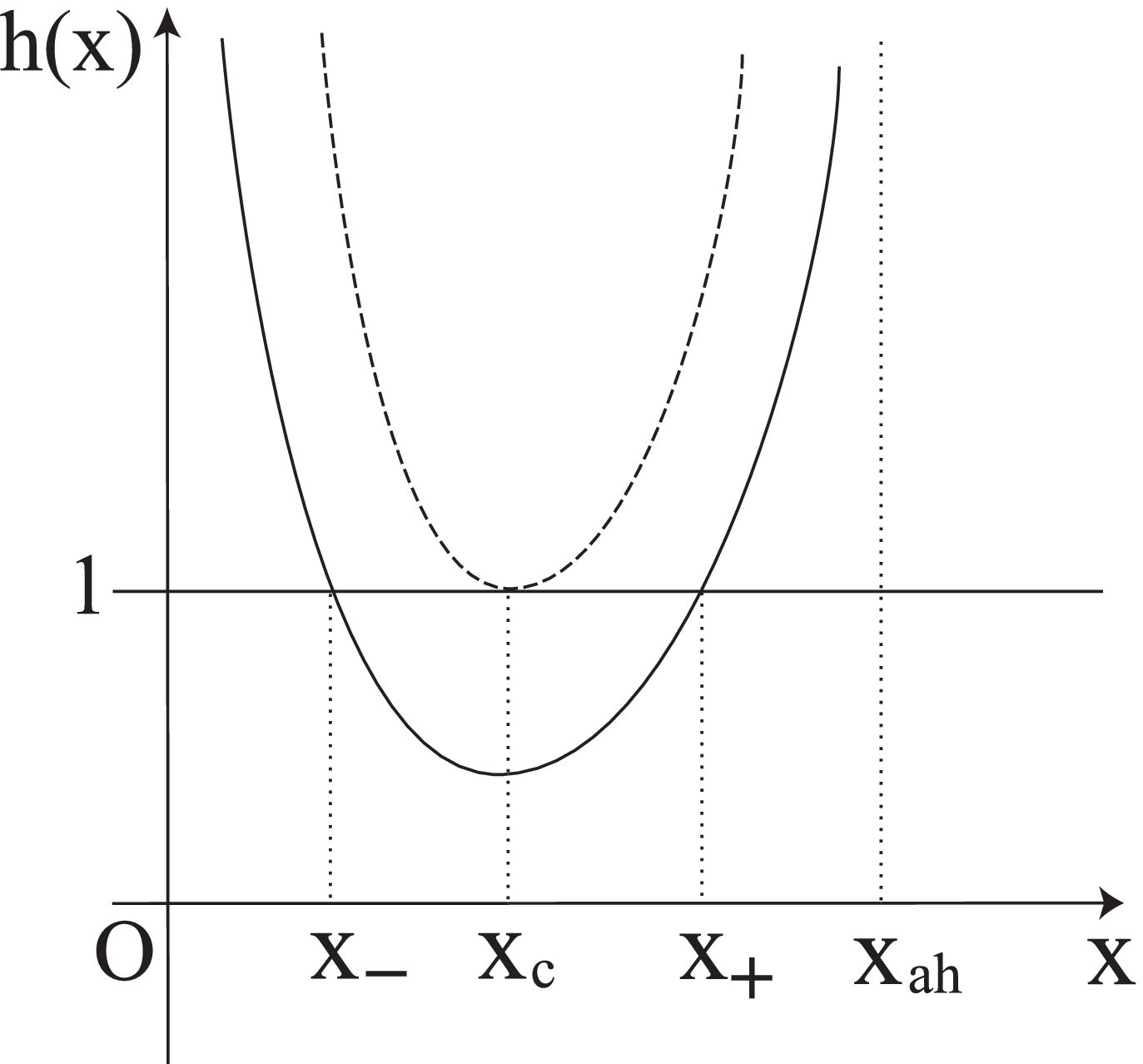} &
			\includegraphics[width=7cm]{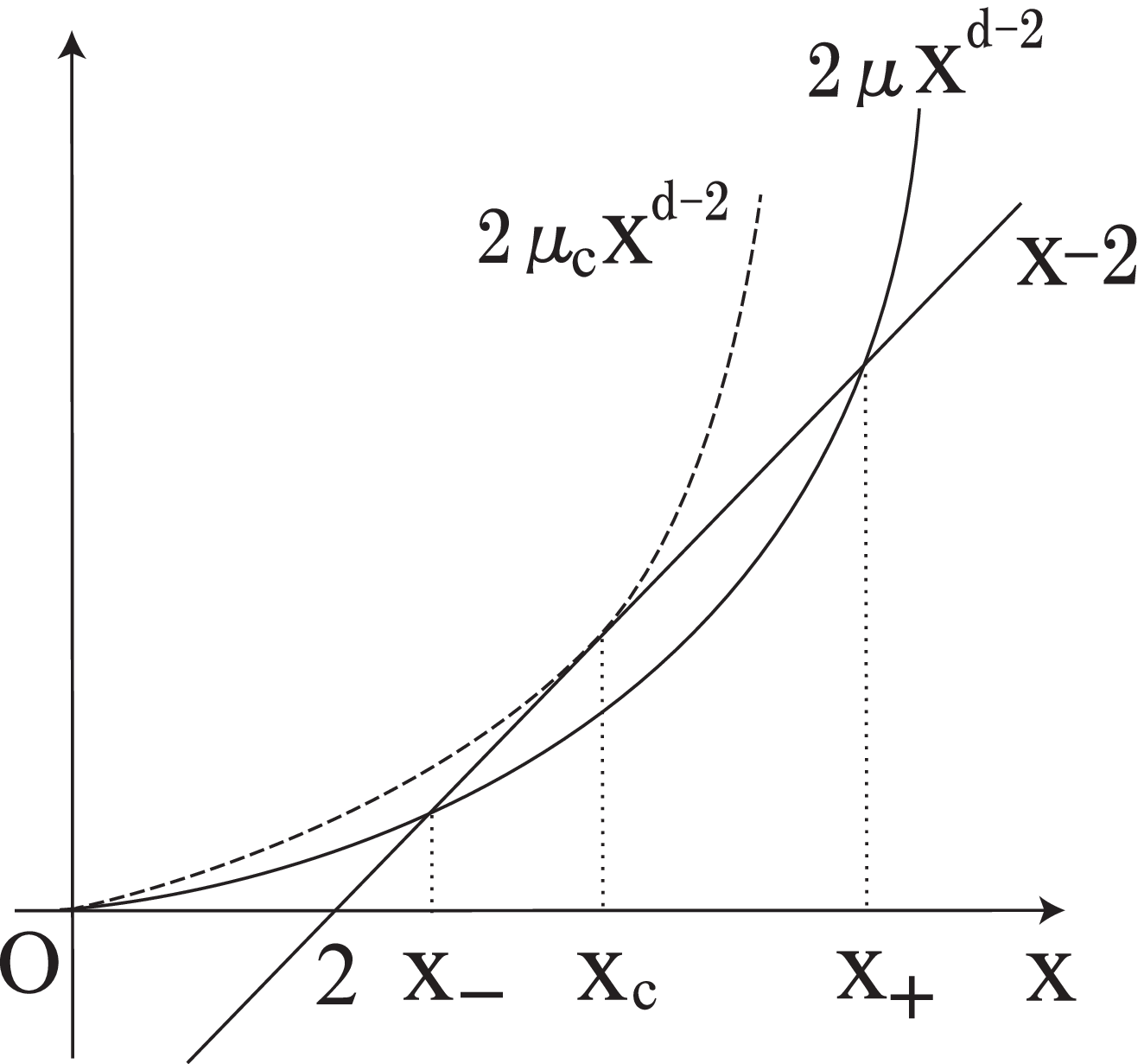} \\
		\end{tabular}
	\caption{{\sf {\footnotesize (a) A schematic graph of $h(x)$ for the generic case $0<\mu<\mu_c$ (solid) and the critical case $\mu=\mu_c$ (dashed). Two roots of $h(x)=1$, denoted by $x_\pm$, determine two null rays $x=x_\pm$. These roots are degenerate ($x_\pm=x_c$) when $\mu=\mu_c$. The apparent horizon is given by $x=x_{\rm ah}$, at which $h(x)$ diverges positively. (b) Schematic graphs of two functions $2\mu x^{d-2}$ and $x-2$, of which intersections determine the two roots. From this picture, the range of two roots \eqref{range} is easily understood. 
}}}
	\label{fg:graph}
	\end{center}
\end{figure}

Instead of constructing double null coordinates in the region II (\eg, according to \cite{Singh:2000sp}), we obtain the trajectory of outgoing radial null rays by solving $ds^2=0$ (with $d\Omega_{d-2}^2=0$), which is written as 
\be
	\frac{dr}{dv}
	=
	\frac{1 - 2\mu x^{d-3}}{2},
\;\;\;
	x := \frac{v}{r}.
\label{drdv}
\ee
What easily seen from this is that the outgoing radial null rays are trapped ($ dr/dv \leq 0 $) in the following region
\be
	x \geq x_{\rm ah} := \frac{1}{( 2\mu )^{1/(d-3)}}.
\ee
Thus, the curve $x=x_{\rm ah}$ gives an apparent horizon. Observe that the intersection of the apparent horizon and the surface of fluid ($v=v_0$) determines a radius $v_0/x_{\rm ah} = (2\mu)^{1/(d-3)} v_0$, which is nothing but the radius of the event horizon $r_{\rm eh}$. In terms of ($x,r$)-coordinates equation \eqref{drdv} is written as
\be
	\frac{dx}{dr}
	=
	\frac{x[h(x)-1]}{r},
\;\;\;
	h(x) := \frac{ 2 }{ x(1-2\mu x^{d-3}) }.
\label{dxdr}
\ee
From this equation, one can easily see that $x=const$, where the constant is a root of $h(x)-1=0$, is an outgoing null ray. One can easily check that the algebraic equation $h(x)-1=0$ is equivalent to 
\be
	q(x) := 2\mu x^{d-2} - (x-2) = 0.
\ee
By a simple algebra one can see that there are two positive roots $x_\pm$ ($x_- \leq x_+$) if and only if the accretion parameter $\mu$ is in the following range
\be
	0
	<
	\mu
	\leq 
	\mu_c
	:=
	\frac{ (d-3)^{d-3} }{ [2(d-2)]^{d-2} }.
\label{mu_range}
\ee
In this case, the roots are in the range of
\be
	2 < x_- \leq x_c \leq x_+,
\;\;\;
	x_c := \frac{2(d-2)}{d-3}.
\label{range}
\ee
When $\mu=\mu_c$ the two roots are degenerate $x_\pm = x_c$. See figures \ref{fg:graph}(a) and \ref{fg:graph}(b), which would be helpful to understand the roots of algebraic equations $h(x)-1=0$ and $q(x)=0$, respectively.

\begin{figure}[bth]
	\begin{center}
		\setlength{\tabcolsep}{ 3 pt }
		\begin{tabular}{ cc }
			(a) & (b) \\
			\includegraphics[width=8cm]{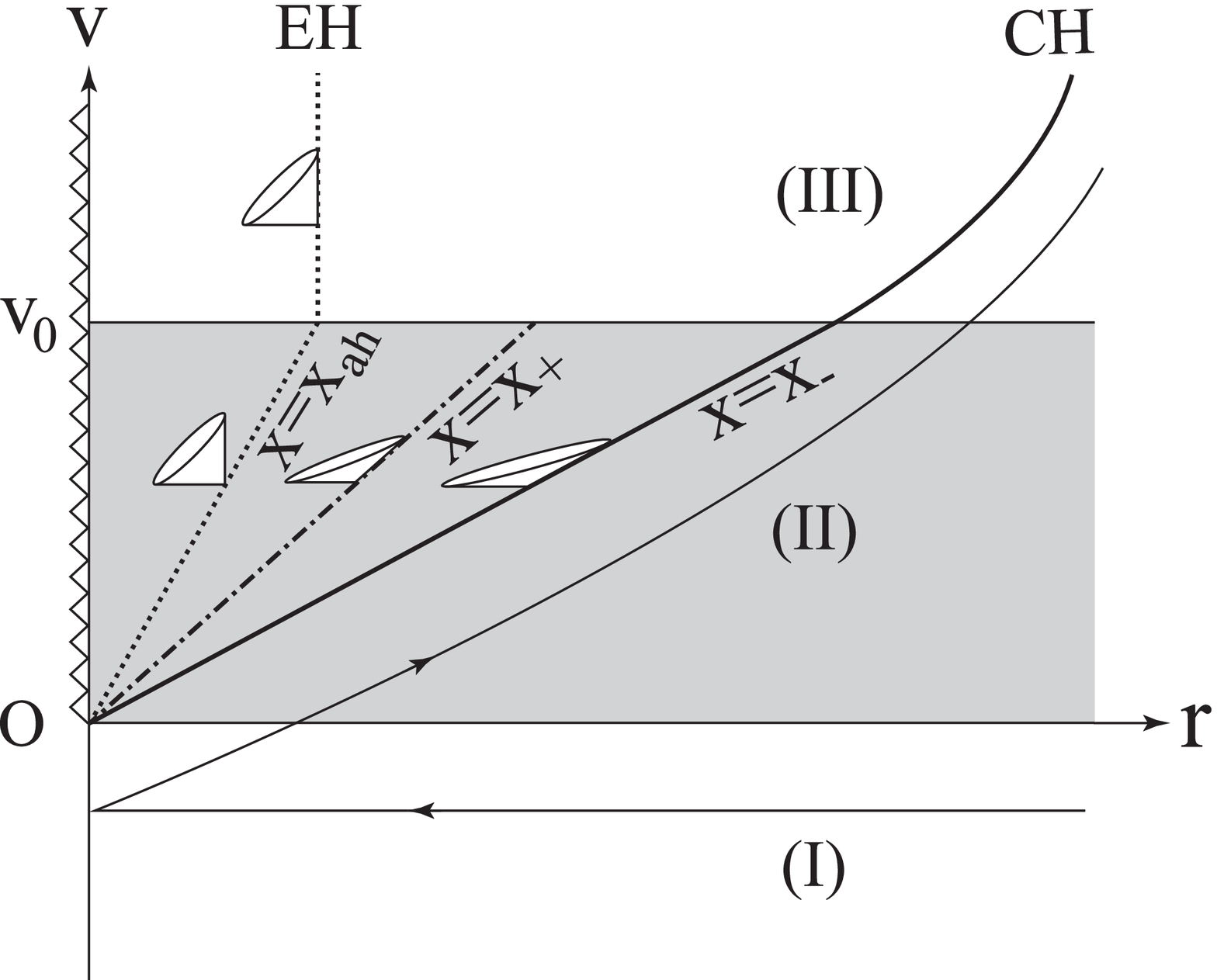} &
			\includegraphics[width=8cm]{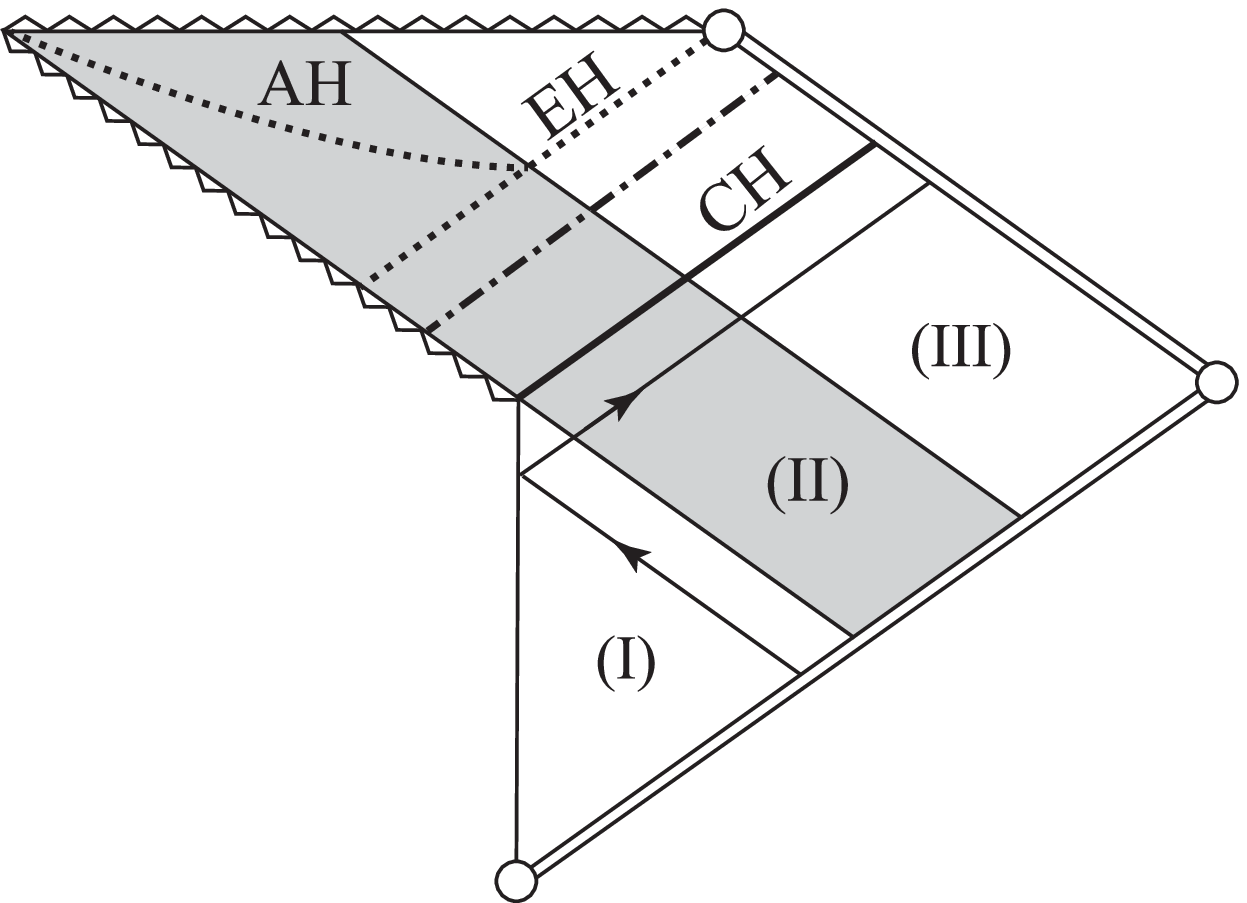} \\
		\end{tabular}
	\caption{{\sf {\footnotesize (a) A schematic spacetime diagram in the $v$-$r$ coordinates. The Cauchy horizon (CH, thick solid) $x=x_-$, the null curve $x=x_+$ (dot-dashed), the apparent horizon (AH, dotted) $x=x_{\rm ah}$, and several future-directed light cones are drawn. In the region II (gray) the null dust fluid infalls toward the center. EH is a part of the event horizon. The curve with two arrows represents a typical null ray that passes through the center just before the appearance of singularity. (b) A corresponding conformal diagram, that manifests the global nakedness of the singularity. When $\mu = \mu_c$ (and in the limit $\mu \to \mu_c$), the CH (thick-solid) and $x=x_+$ (dot-dashed) are coincide, but the key causal structures do not change. Namely, the singularity is still globally naked, rather than marginally naked.
}}}
	\label{fg:diagram}
	\end{center}
\end{figure}

By simple arguments, one can show that the singularity located at $r=0$ and $v>0$ is spacelike, whereas the singularity located at $(v,r)=(0,0)$ is a globally naked one for $\mu$ being in the range \eqref{mu_range} (see, \eg, \cite{JoshiBook}). In particular, the null ray $x=x_-$ is the first outgoing null ray emanating from the singularity (see, \eg, appendix A in \cite{Miyamoto:2003wr}). Namely, the $x=x_-$ gives (a part of) the Cauchy horizon. See figures \ref{fg:diagram}(a) and \ref{fg:diagram}(b) for a schematic ($v,r$)-diagram and a conformal diagram, respectively.\footnote{We stress that the null ray $x=x_+$ is an event horizon {\it if} the positive $v$-region is filled entirely with the null dust. Since we cut the Vaidya region and connect it to the outer vacuum region, the null ray $x=x_+$ plays no special role in the present analysis. Accordingly, neither the limit $\mu \to \mu_c$ nor exactly $\mu=\mu_c$ case correspond to a marginally naked singularity.}

%----------------------------------------------------------------------%
%----------------------------------------------------------------------%
%----------------------------------------------------------------------%
%----------------------------------------------------------------------%

%----------------------------------------------------------------------%
%----------------------------------------------------------------------%
%----------------------------------------------------------------------%
%----------------------------------------------------------------------%

%----------------------------------------------------------------------%
%----------------------------------------------------------------------%
\section{Null geodesics near the Cauchy horizon}
\label{sec:null}
%----------------------------------------------------------------------%
%----------------------------------------------------------------------%

Integrating equation \eqref{dxdr}, we obtain a formal expression of the outgoing radial null ray
\be
	\frac{r}{r_0}
	=
	\exp \left[
		\int_0^x H(x) dx
	\right],
\;\;\;
	H(x) := \frac{1}{x[h(x)-1]}
	=
	- \frac{ 2\mu x^{d-3}-1 }{ q(x) }.
\label{r/r0}
\ee
Here, $r_0$ is a constant corresponding to the radius when the outgoing null ray passes the $r$-axis ($v=0, r>0$. Namely $x=0$).

We are interested in the null rays passing near the naked singularity located at $(v,r)=(0,0)$, which are responsible for the particle creation. Since the Cauchy horizon is given by $x=x_-$, we have to evaluate the integral in equation \eqref{r/r0} near the pole of the integrand where $q(x)=0$. In the following, we evaluate the integral near the Cauchy horizon for the non-degenerate case ($0<\mu<\mu_c$) and the degenerate case ($\mu=\mu_c$) separately. We mostly use the notation and techniques developed in \cite{Miyamoto:2003wr}.

%----------------------------------------------------------------------%
\subsection{Generic case ($0<\mu<\mu_c$)}
%----------------------------------------------------------------------%

For $0<\mu<\mu_c$, the algebraic equation $h(x)=1$ has a simple root at $x_-$. Therefore, we subtract a pole from the integrand in equation \eqref{r/r0} as
\begin{align}
\begin{split}
	\frac{r}{r_0}
	&=
	\exp \left[
		\int_0^x
		\frac{1}{\gamma(x-x_-)}
		dx
	\right]
	\exp \left[
		\int_0^x
		\left(
		H(x) - \frac{1}{\gamma(x-x_-)}
		\right)
		dx
	\right]
\\
	&=
	\left( \frac{x_- - x}{x_-} \right)^{1/\gamma}
	\exp \left[
		\int_0^x
		H_\ast (x)
		dx
	\right],
\end{split}
\label{r/r0_generic}
\end{align}
where
\be
	H_\ast (x) := H(x) - \frac{1}{\gamma(x-x_-)},
\;\;\;
	\gamma := x_- h^\prime (x_-)
	=
	- \frac12 (d-3)(x_c-x_-).
\label{gamma}
\ee
Note that function $H_\ast(x)$ converges to a certain constant in the limit $x \to x_- -0$.

We are ready to obtain the map of null rays that gives the relation $v=G(u)$ between the advance time $v$ at which an ingoing null ray departs from the past null infinity and the retarded time $u$ at which this null ray terminates at the future null infinity after passing through the regular center (\ie, $r=0, v<0$). Suppose an ingoing null ray $v = v_{\rm in} = const \; (< 0)$ propagating in the region I (see figure \ref{fg:diagram}). This null ray turns to an outgoing null ray $\bar{u} = v_{\rm in} $ after passing through the regular center. This null ray passes through the I-II boundary ($v = 0, r>0$, \ie, $x=0$) and is expressed by \eqref{r/r0_generic} with the integration constant $r_0$ given by
\be
	r_0 = -\frac{v_{\rm in}}{2}.
\label{r0}
\ee
Then, this null ray reaches the II-III boundary ($v=v_0$) and is expressed by $u=u_{\rm out}=const$. This constant $u_{\rm out}$ is given by the right-hand side of equation \eqref{U} with $v=v_0$ and
\be
	r
	=
	\frac{ v_0 }{ x }
	=
	\frac{v_0 }{ x_-}
	+
	\frac{ v_0  }{ x_-^2 }(x_- -x ) + O\left( (x_- - x)^2 \right).
\label{r_leading}
\ee
Namely, substituting the above expression of $r$ into equation \eqref{U}, we have at the leading order
\be
	u_{\rm out}
	=
	u_0 
	-
	\frac{2 r_- }{x_- f( r_- )} (x_- - x)
	+
	O\left( (x_- - x)^2 \right),
\;\;\;
	r_- := \frac{ v_0 }{ x_- },
\label{Uout}
\ee
where $ u = u_0 := v_0 - 2 r_\ast |_{r=v_0/x_-}$ gives the Cauchy horizon in the region III.

On the other hand, substituting equations \eqref{r0} and \eqref{r_leading} into \eqref{r/r0_generic}, we obtain
\be
	\frac{ x_- - x }{ x_- }
	\simeq
	\left( \frac{2r_-}{ I } \right)^\gamma ( -v_{\rm in} )^{-\gamma},
\;\;\;
	I := \exp \left[ \int_0^{x_-} H_\ast (x) dx \right]
\label{deltax_generic}
\ee
up to $O\left( (x_- - x)^2 \right)$.

Eliminating $(x_- - x)$ from equations \eqref{Uout} and \eqref{deltax_generic}, and omitting the subscripts `in' and `out', we obtain the map of null ray $v=G(u)$ as,\footnote{Let us point out a typo in reference \cite{Singh:2000sp}, in which the four-dimensional case was analyzed. For $d=4$ the explicit expression of the two roots are $x_\pm = (1\pm\sqrt{1-16\mu})/4\mu$. With using this and equations \eqref{gamma} and \eqref{G} one obtains
\be
	\alpha
	=
	-\frac{1}{\gamma}
	=
	\frac{ 1+\sqrt{1-16\mu} }{ 2\sqrt{ 1-16\mu } }.
\ee
This quantity $\alpha$ should be identical with $A_-$ in \cite{Singh:2000sp} [see equation (16)]. However, the expression of $A_\pm$ in \cite{Singh:2000sp} seems incorrect. A correct expression should be
\be
	A_\pm
	=
	\mp \frac{ 1-2\mu \alpha_\pm }{ 2\mu ( \alpha_+ - \alpha_- ) }, 
\ee
where $\alpha_\pm$ is $x_\pm$ in our notation. With this corrected expression, one can show $\alpha = A_-$.}
\be
	G(u)
	=
	- \frac{ f^\alpha (r_-) }{ (2r_-)^{\alpha-1} I }
	( u_0 -u )^\alpha,
\;\;\;
	\alpha
	:=
	- \frac{1}{\gamma}
	=
	\frac{ 2 }{ (d-3)(x_c-x_-) }.
\label{G}
\ee

%----------------------------------------------------------------------%
\subsection{Degenerate case ($\mu=\mu_c$)}
%----------------------------------------------------------------------%

Now, let us consider the critical case in which the algebraic equation $q(x)=0$ has a degenerate root at $x=x_\pm = x_c$. In a similar way to the generic case, we subtract a pole of the integrand in equation \eqref{r/r0} as
\begin{align}
\begin{split}
	\frac{r}{r_0}
	&=
	\exp \left[
		\int_0^x
		\frac{1}{\hat{\gamma}(x-x_c)^2}
		dx
	\right]
	\exp \left[
		\int_0^x
		\left(
		H(x) - \frac{1}{\hat{\gamma}(x-x_c)^2}
		\right)
		dx
	\right]
\\
	&=
	\exp \left[ \frac{1}{\hat{\gamma}(x_c -x)} \right]
	\exp \left[
		- \frac{1}{\hat{\gamma} x_c} + 
		\int_0^x
		\hat{H}_\ast (x)
		dx
	\right],
\end{split}
\label{r/r0_deg}
\end{align}
where
\be
	\hat{H}_\ast (x) := H(x) - \frac{1}{\hat{\gamma}(x-x_c)^2},
\;\;\;
	\hat{\gamma} := \frac12 x_c h^{\prime\prime} (x_c)
	=
	\frac{d-3}{4}.
\label{gamma_crit}
\ee
Note that $\hat{H}_\ast(x)$ is regular at the Cauchy horizon $x=x_c$ and its integration is finite in the limit $x \to x_c - 0$.

A counterpart of equation \eqref{deltax_generic} in the present case is given by
\be
	(x_c - x)^{-1}
	\simeq
	\hat{\gamma} \ln \left[ \frac{ 2r_c }{ \hat{I} } ( -v_{\rm in} )^{-1}  \right],
\label{deltax_crit}
\ee
where
\be
	r_c := \frac{  v_0 }{  x_c },
\;\;\;
	\hat{I} :=
		\exp \left[
		- \frac{1}{\hat{\gamma} x_c} + 
		\int_0^{x_c}
		\hat{H}_\ast (x)
		dx
	\right].
\ee
Eliminating ($x_c-x$) from equations \eqref{Uout} and \eqref{deltax_crit}, and omitting the subscripts `in' and `out', we obtain the map of null ray $v=G(u)$ in the case where the Cauchy horizon is degenerate,
\be
	G(u)
	=
	-\frac{2r_c}{ \hat{I} }
	\exp\left[
		- \frac{4 r_c }{ (d-2) f(r_c) ( u_0-u )  }
	\right].
\label{G_crit}
\ee

%----------------------------------------------------------------------%
%----------------------------------------------------------------------%
\section{Power and energy of particle creation}
\label{sec:power}
%----------------------------------------------------------------------%
%----------------------------------------------------------------------%

Now, we are ready to evaluate the particle creation under the geometric-optics approximation. We consider a massless scalar field $\phi$ coupled to the Ricci scalar curvature $R$ as
\be
	(\dalm - \xi R) \phi = 0,
\label{eom}
\ee
where $\xi$ is an arbitrary (real) constant. In particular, the cases of $\xi = 0$ and $\xi = \xi_d := (d-2)/[4(d-1)] $ are called the minimal coupling and conformal coupling, respectively (see appendix \ref{sec:conformal}).

We assume as usual that the quantum state is in the vacuum in which positive-energy ingoing particles are absent at the past null infinity. Then, the collapsing spacetime excites the quantum field, and one can expect a positive-energy flux is observed in the asymptotic region. The power $P$ (the energy emitted per unit time) is given by the integration of the vacuum expectation value of stress-energy tensor over the ($d-2$)-sphere in the late-time asymptotic region. The formula obtained with the geometric-optics approximation and the point-splitting regularization (see appendix \ref{sec:quantum}) is
\be
	P(u)
	=
	\frac{ 1 }{4\pi} 
	\left[
		\left( \frac14 - \xi \right) 
		\left(
			\frac{G^{\prime\prime}(u)}{G^\prime(u)}
		\right)^2
		+
		\left( \xi - \frac16 \right)
		\frac{ G^{\prime\prime\prime}(u) }{ G^\prime(u) }
	\right],
\label{power}
\ee
where $G(u)$ is the map obtained in the previous section. The total energy radiated is the integration of this power by the retarded time,
\be
	E(u)
	=
	\int_{-\infty}^u P(u) du.
\label{energy}
\ee

As shown in appendix \ref{sec:quantum} the actual formula is given by the sum of the power (and energy) given here over all $l$ (\ie, angular quantum numbers). Since the power and energy given here are independent of $l$, those sum diverge. Such a divergence is due to the fact that we ignore the back scattering by potential barriers, which certainly will reduce the emission by highly rotational modes. Hereafter, we omit the sum over $l$, and it should be simply kept in mind that the above formulae take into account only the small-$l$ modes.

%----------------------------------------------------------------------%
\subsection{Generic case ($0<\mu<\mu_c$)}
%----------------------------------------------------------------------%

Substituting the map of null rays for $0<\mu<\mu_c$, equation \eqref{G}, into formulae \eqref{power} and \eqref{energy}, we obtain 
\begin{align}
\begin{split}
	P
	&=
	\frac{ (\alpha-1)(\alpha+1-12\xi) }{ 48\pi } (u_0-u)^{-2},
\\
	E
	&=
	\frac{ (\alpha-1)(\alpha+1-12\xi) }{ 48\pi } (u_0-u)^{-1}.
\label{power_generic}
\end{split}
\end{align}
Thus, we reproduce and generalize to general dimensions the result in \cite{Singh:2000sp,Miyamoto:2003wr} that the power diverges as the quadratic inverse of the remaining time to the Cauchy horizon $(u_0-u)$.

The factor in the power and energy in equation \eqref{power_generic}, $A:=(\alpha-1)(\alpha+1-12\xi)$, depends on $\alpha$, which is a function of accretion parameter $\mu$, and the coupling constant $\xi$. Although we have no explicit expression of $\alpha = \alpha(\mu)$ for general $d$ except for $d=4$, we can discuss the ($\mu,\xi$)-dependence of $A$ in general by observing the following facts. With using equation \eqref{G} one can easily obtain
\be
	\alpha-1 = \frac{ x_- - 2 }{ x_c - x_- }.
\label{alpha-1}
\ee
Taking into account this equation and the range of $x_-$ and $x_c$ given in equation \eqref{range} one can easily show that $\lim_{\mu \to 0}\alpha(\mu) = 1$ and $\lim_{\mu \to \mu_c} \alpha(\mu) = +\infty$. Furthermore, one can show that $\alpha(\mu)$ is an increasing function ($d\alpha/d\mu > 0$) from that $x_-$ is an increasing function of $\mu$, as obvious from figure \ref{fg:graph}(b). Thus, we have $\alpha(\mu)>1$ in general ($0<\mu<\mu_c$) and $\alpha$ diverges positively in the limit $\mu \to \mu_c$. In appendix \ref{sec:redshift} it is shown that this inequality $\alpha>1$ is equivalent to that the redshift of outgoing null rays diverges at the Cauchy horizon.

From the above observations of $\alpha(\mu)$ one finds several properties of the power and energy. The factor $A$ diverges in the limit $\mu \to \mu_c$ for any finite coupling constant $\xi$. For other generic case of $0<\mu<\mu_c$, the factor $A$ is positive definite if the coupling is `weak,' $ \xi < (\alpha+1)/12$. Note that this case includes the minimal coupling $\xi=0$ as a special case, where $A = \alpha^2-1 $ holds. On the other hand, the factor $A$ is non-positive if the coupling is `strong,' $\xi \geq (\alpha+1)/12$. We should stress that the conformal coupling $\xi = \xi_d$ plays no special role in general dimensions except for the four-dimensional case, in which $A=(\alpha-1)^2 >0$ holds.

%----------------------------------------------------------------------%
\subsection{Degenerate case ($\mu = \mu_c$)}
%----------------------------------------------------------------------%

Substituting the map of null rays for $\mu=\mu_c$, equation \eqref{G_crit}, into the formulae \eqref{power} and \eqref{energy}, we obtain the power and energy at the leading order,
\begin{align}
\begin{split}
	P
	&=
	\frac{ r_c^2 }{ 3\pi (d-2)^2 f^2(r_c) } (u_0 - u)^{-4},
\\
	E
	&=
	\frac{ r_c^2 }{ 9\pi (d-2)^2 f^2(r_c) } (u_0 - u)^{-3}.
\label{power_crit}
\end{split}
\end{align}
Namely, the power (resp.\ energy) diverges as the quartic (cubic) inverse of the remaining time to the Cauchy horizon.
These results have not been known even in the four-dimensional case and are obtained for the first time. It is quite interesting to notice that according to the quartic and cubic behaviors, a scale determined by the background $r_c := v_0 / x_c $ enters into equation \eqref{power_crit},\footnote{Note that $f(r_c)$ appearing in \eqref{power_crit} is just a number: $ f(r_c) = (d-3)/(d-2) $.} in contrast to the generic case discussed in the previous subsection. This quantity $r_c$ scales with the total mass of collapsing fluid $M$ as $r_c \sim M^{1/(d-3)}$. This means that the behaviors of power and energy cannot be predicted only on a dimensional basis in spite of the scale invariance of the central self-similar region. We should stress also that the cancellation of coupling constant $\xi$ has happened and the final results \eqref{power_crit} are independent of $\xi$. 

%----------------------------------------------------------------------%
%----------------------------------------------------------------------%
\section{Discussions}
\label{sec:conc}
%----------------------------------------------------------------------%
%----------------------------------------------------------------------%

Motivated by the recent argument that the trans-Planckian domains of spacetime not veiled by horizons, called the visible border of spacetime, will be generated by high-energy particle collisions in the context of TeV-scale gravity, we have investigated the particle creation by the naked-singularity formation in general dimension, which possibly plays important roles in collider experiments. While the actual generation will be highly asymmetric phenomena, we have assumed just for simplicity that the background is perfectly spherically symmetric and modeled by the self-similar Vaidya solution \eqref{metric}, describing the collapse of the pressureless lightlike fluid. As the results, we have obtained the formulae of emission power and energy, equation \eqref{power_generic} for the generic case ($0<\mu<\mu_c$) and equation \eqref{power_crit} for the critical case ($\mu=\mu_c$), where $\mu$ is a dimensionless accretion parameter of the fluid \eqref{m}.

In the latter case ($\mu=\mu_c$) the Cauchy horizon is `degenerate', and the resultant formula has not been known even in the four-dimensional case. Although this case is just a particular point in the parameter space, the result is somewhat interesting at a theoretical level in the sense that the power depends on the background dimensionful parameter $r_c \sim M^{1/(d-3)}$ despite the scale invariance of the central region. Incidentally, the present authors confess that they have no clear explanation why the behaviors of the power and energy are so different between the limit of $\mu \to \mu_c$ and the case of exactly $\mu=\mu_c$.

We comment on the validity of approximations adopted for simplicity in this paper. The actual visible-border production is expected to be a highly dynamical process, of which a typical time scale may be given by the light-crossing time through the colliding region. Therefore, the validity of the geometric-optics approximation and/or the quantum field theory in classical background itself could be questionable. This point would be worth further considerations. Incidentally, we should mention the difficulty to verify the geometric-optics approximation in naked-singularity formation in general,\footnote{The geometric-optics approximation can be an exact method in two dimensions \cite{Birrell:1982ix}.} in contrast to the black-hole formation. As shown in appendix \ref{sec:redshift}, the redshift of the outgoing waves diverges at the Cauchy horizon. Thus, the wavelength of the particles detected in the asymptotic region becomes (possibly, quite) short (\ie, blueshifted) in the central region. However, this does not imply the validity of the geometric-optics approximation necessarily since the curvature around the singularity is arbitrarily large. Furthermore, remember that there is the fundamental problem that the spectrum of created particle cannot be calculated uniquely due to the ambiguity of boundary conditions at the singularity. Therefore, one cannot know the typical energy of particles detected in the asymptotic region, and therefore cannot know the energy of particles propagated back to the region around the singularity.

The divergence of the energy emitted seems to suggest that the backreaction to the geometry should be taken into account in an actual dynamics. In other words, the divergence suggests the existence of a ``semiclassical instability.'' We should mention reference \cite{Harada:2000me} here, however, in which Harada \et\ argued that if  a quantum gravitational effect works as a cutoff of the radiation, the total energy radiated only amounts to a few amounts of Planck energy, which means that the backreaction is negligible in the collapse of stellar-size massive objects. On the other hand, if we repeat their argument \cite{Harada:2000me} in the present TeV-gravity context, {\it both} the energy of background and the net energy radiated are of the order of TeV. Namely, if a quantum gravitational cutoff is switched on, say, when the remaining time is the Planck time $u_0-u \sim t_P$, the net energy radiated by this moment amounts to $E \sim M_P c^2$ from equation \eqref{power_generic} for the generic case and $E \sim M_P c^2 ( M/M_P )^{2/(d-3)} $ from equation \eqref{power_crit} for the degenerate case. Thus, we naturally expect that the backreaction will modify the dynamics. This difference of the significance of semiclassical effects between the stellar collapse in general relativity and the particle collisions in TeV-gravity is worth being stressed.

%----------------------------------------------------------------------%
%----------------------------------------------------------------------%
\subsection*{Acknowledgments}
%----------------------------------------------------------------------%
%----------------------------------------------------------------------%

UM would like to thank Tomohiro Harada for useful discussions. UM is supported by Research Center for Measurement in Advanced Science in Rikkyo University, and by the Grant-in-Aid for Scientific Research Fund of the Ministry of Education, Culture, Sports, Science and Technology, Japan [Young Scientists (B) 22740176].

\appendix
%----------------------------------------------------------------------%
%----------------------------------------------------------------------%
\section{Conformal coupling}
\label{sec:conformal}
%----------------------------------------------------------------------%
%----------------------------------------------------------------------%

An action of the scalar field $\phi$ that couples to the Ricci scalar curvature $R$ may be given by
\be
	S[\phi] =
	\int d^d x \sqrt{-g}
	\left(
		-\frac12 (\nb \phi)^2 -\frac12 \xi R \phi^2
	\right),
\ee
where $\xi$ is a coupling constant. The energy-momentum tensor derived from this action is
\be
	T_{\mu\nu}
	=
	\nb_\mu \phi \nb_\nu \phi - \frac12 g_{\mu\nu} (\nb \phi)^2
	+
	\xi \left(
		G_{\mu\nu} \phi^2 - \nb_\mu \nb_\nu \phi^2 + g_{\mu\nu} \dalm \phi^2
	\right),
\ee
whereas the equation of motion is given by \eqref{eom}.

Let us consider a conformal transformation $ g_{\mu\nu} \to \bar{g}_{\mu\nu} = e^{2\omega} g_{\mu\nu} $, where $\omega(x)$ is an arbitrary scalar function. It is noted that under this transformation the d'Alembertian (operating on a scalar field $\psi$) and the Ricci scalar curvature transform as
\begin{align}
\begin{split}
	\bar{\dalm} \psi
	&=
	e^{-2\omega}
	\left[
		\dalm \psi 
		+
		(d-2) \nb \psi \cdot \nb \omega
	\right],
\\
	\bar{R}
	&=
	e^{-2\omega}
	\left[
		R - 2(d-1) \dalm \omega - (d-2)(d-1) ( \nb \omega )^2
	\right].
\end{split}
\label{conf_dalm}
\end{align}
Assuming that the scalar field transforms as $\phi \to \bar{\phi} = e^{a \omega} \phi$ with a constant $a$, the equation of motion \eqref{eom} transforms as
\begin{multline}
	( \bar{\dalm} -\xi \bar{R} ) \bar{\phi}
	=
	e^{(a-2)\omega}
	\Big(
		\dalm \phi - \xi R \phi
		+
		[ a + 2(d-1)\xi ] \phi \dalm \omega
\\
		+
		[ a (a +d-2) + (d-2)(d-1)\xi ] \phi ( \nb \omega )^2
		+
		( 2a + d-2 ) \nb \phi \cdot \nb \omega
	\Big).
\end{multline}
Therefore, if one chooses the coupling constant $\xi$ and $a$ as
\be
	\xi = \xi_d := \frac{d-2}{4(d-1)},
\;\;\;
	a = - \frac{d-2}{2},
\ee
the equation of motion \eqref{eom} is invariant under the conformal transformation. Namely,
\be
	\left(
		\bar{ \dalm }
		-
		\frac{d-2}{4(d-1)} \bar{R}
	\right) \bar{\phi}
	=
	e^{-(d+2)\omega/2}
	\left(
		\dalm - \frac{d-2}{4(d-1)} R
	\right) \phi,
\;\;\;
	\bar{\phi} :=  e^{-(d-2)\omega/2} \phi.
\ee

%----------------------------------------------------------------------%
%----------------------------------------------------------------------%
\section{Quantization}
\label{sec:quantum}
%----------------------------------------------------------------------%
%----------------------------------------------------------------------%

Here, we formulate the quantization of a scalar field in $d$-dimensions that couples to the scalar curvature in the manner described above. In particular, we derive the formula of emission power, generalizing the results in \cite{Ford:1978ip,Miyamoto:2004ba} to arbitrary dimensions and the generally coupling scalar field. 

In the asymptotic region ($ r \to \infty $) a mode function of the scalar field obeying equation of motion \eqref{eom} is given by
\be
	p_{\omega l}
	\simeq
	\frac{1}{\sqrt{ 4\pi \omega  } \; r^{(d-2)/2} }
	\left(
		e^{-i\omega v} + e^{ -i\omega G(u) }
	\right) Y_l (\Omega),
\;\;\;
	\omega > 0.
\label{mode}
\ee
Here, $Y_l (\Omega)$ is a normalized scalar harmonics on the ($d-2$)-sphere
\be
	\left[ \Delta_{d-2} + l(l+d-3) \right] Y_l (\Omega) = 0,
\;\;\;
	l = 0,1,2,\ldots ,
\ee
where $\Delta_{d-2}$ is the Laplacian on the sphere. $u \simeq t-r$ and $v \simeq t+r$ are the retarded and advanced time coordinates, respectively, in the quasi-Minkowski region. We note that quantum numbers associated to the other angular degrees of freedom are omitted. The mode function \eqref{mode} behaves as an ingoing wave at the past null infinity, whereas it behaves as a redshifted outgoing wave at the future null infinity. When the spacetime is globally flat, $G(u)=u$ holds. The normalization constant is chosen so that the mode function is normalized as
\be
	( p_{\omega l}, p_{\omega^\prime l^\prime} ) = \delta(\omega-\omega^\prime) \delta_{ll^\prime},
\ee
where $(\cdot, \cdot)$ denotes the Klein-Goldon inner product defined by
\be
	(p_1,p_2)
	:=
	-i \int_\Sigma ( p_1 \pd_\mu p_2^\ast - p_2^\ast \pd_\mu p_1 ) \sqrt{g_\Sigma} d\Sigma^\mu.
\ee
Here, $\Sigma$ is a spacelike hypersurface with the volume element $ \sqrt{g_\Sigma} d\Sigma^\mu$.

The field operator can be expanded by the above mode function as
\be
	\phi
	=
	\sum_l \int_0^\infty d\omega
	\left(
		a_{\omega l} p_{\omega l} + a_{\omega l}^\dagger p_{\omega l}^\ast
	\right)
\label{expand}
\ee
with the annihilation operator $a_{\omega l}$ and creation operator $a_{\omega l}^\dagger$ satisfying the usual commutation relation,
\be
	[a_{\omega l},a^\dagger_{\omega^\prime l^\prime}] = \delta(\omega-\omega^\prime) \delta_{ll^\prime}.
\ee
The quantum field is assumed to be in the vacuum $ |0 \rangle $ (eternally, since we work in the Heisenberg picture) defined by
\be
	a_{\omega l} |0 \rangle = 0,
\;\;\;
	\mbox{for all}
\;\;\;
	\omega, l.
\ee

The power is the vacuum expectation value (VEV) of the following ($t,r$)-component of the energy-momentum tensor at future null infinity
\be
	T^t_r
	\simeq
	-\frac12
	\left(
		\phi_{,r} \phi_{,t} + \phi_{,r} \phi_{,t}
	\right)
	+
	\xi
	\left(
		\phi \phi_{,r} + \phi_{,r} \phi
	\right)_{,t} ,
\label{Ttr}
\ee
where we have symmetrized the products. Substituting the expansion \eqref{expand} into this equation, we obtain
\be
	\langle 0 | T^t_r | 0 \rangle
	=
	\sum_l \int_0^\infty d\omega
	 \left[
		-\frac12
		\left(
			p_{\omega l, t} p^\ast_{\omega l, r} +  p_{\omega l, r} p^\ast_{\omega l, t}
		\right)
		+ \xi
		\left(
			p_{\omega l} p^\ast_{\omega l, r} + p_{\omega l, r} p^\ast_{\omega l}
		\right)_{,t}
	\right].
\label{vev}
\ee
Here, according to \cite{Ford:1978ip} we prescribe the point-splitting regularization scheme to this integration. Namely, in order to regulate the divergence of integral due to the simultaneous evaluation at a point, we displace the arguments of $p_{\omega l}^\ast$ in equation \eqref{vev} as $ (u,v) \to (u+\epsilon,v+\epsilon) $ with an infinitesimal distance $\epsilon$. Then, such a prescribed VEV reads
\begin{multline}
	\langle 0 | T^t_r | 0 \rangle_\epsilon
	=
	\frac{ \sum_{l} | Y_l |^2 }{4\pi r^{d-2}} 
	\int_0^\infty d\omega
	\Big[
		G^\prime(u)G^\prime(u+\epsilon) \omega e^{i\omega[ G(u+\epsilon)-G(u) ]} - \omega e^{i\omega\epsilon}
		- i\xi
		\Big(
			[ G^\prime(u+\epsilon)+1 ]e^{i\omega[ G(u+\epsilon)-v ]}
\\
			-
			[ G^\prime(u)+1 ] e^{ -i\omega[ G(u)-v-\epsilon ] }
			+
			[ G^\prime(u+\epsilon)-G^\prime(u) ] e^{i\omega[G(u+\epsilon)-G(u)]}
		\Big)_{,t}
	\Big].
\end{multline}
Implementing the integrations over $\omega$, one obtains
\begin{align}
	\langle 0 | T^t_r | 0 \rangle_\epsilon
	&=
	\frac{ \sum_{l} | Y_l |^2 }{4\pi r^{d-2}} 
	\left[
		- \frac{ G^\prime(u)G^\prime(u+\epsilon) }{ [ G(u+\epsilon)-G(u) ]^2 }
		+ \frac{1}{\epsilon^2}
		+ \xi
		\left(
			\frac{ G^\prime(u+\epsilon)-G^\prime(u) }{ G(u+\epsilon)-G(u) }
		\right)_{,u} + O(\epsilon)
	\right]
\nn
\\
	&=
	\frac{ \sum_{l} | Y_l |^2 }{4\pi r^{d-2}} 
	\left[
		\left( \frac14 - \xi \right) 
		\left(
			\frac{G^{\prime\prime}(u)}{G^\prime(u)}
		\right)^2
		+
		\left( \xi - \frac16 \right)
		\frac{ G^{\prime\prime\prime}(u) }{ G^\prime(u) }
	\right] + O(\epsilon).
\end{align}
Note that the singular term $\epsilon^{-2}$ disappears and the leading-order is $O(\epsilon^0)$ in the final expression. The power is defined by the integrating of $\lim_{\epsilon \to 0} \langle 0 | T^t_r | 0 \rangle_\epsilon $ over the $(d-2)$-sphere of a large radius $r$,
\be
	P(u)
	:=
	\int \langle 0 | T^t_r | 0 \rangle r^{d-2} d\Omega_{d-2}
	=
	\sum_l \frac{ 1 }{4\pi} 
	\left[
		\left( \frac14 - \xi \right) 
		\left(
			\frac{G^{\prime\prime}(u)}{G^\prime(u)}
		\right)^2
		+
		\left( \xi - \frac16 \right)
		\frac{ G^{\prime\prime\prime}(u) }{ G^\prime(u) }
	\right].
\ee

%----------------------------------------------------------------------%
%----------------------------------------------------------------------%	
\section{Redshift}
\label{sec:redshift}
%----------------------------------------------------------------------%
%----------------------------------------------------------------------%

The tangent of a null geodesic $k^\mu := dx^\mu/d\lambda$, where $\lambda$ is an affine parameter, is obtained by solving $k^\mu \nb_\mu k^\nu = 0$. In the ($v,r$)-coordinates the $v$-component (\ie, the frequency) of such an equation for a radial null geodesic is\footnote{Only non-vanishing component of the Levi-Civita connection involved is $\Gamma^v_{vv} =\mu(d-3)x^{d-3}/r $.}
\be
	\frac{ d k^v }{ d \lambda } + \frac{ \mu(d-3) x^{d-3} }{ r } ( k^v )^2 = 0.
\label{tangent}
\ee
From the null condition $k^\mu k_\mu = 0$, we have
\be
	k^r = \frac12 ( 1-2\mu x^{d-3} ) k^v.
\ee
With using this relation, the following holds for the derivative of a function of $x$,
\be
	\frac{ d }{ d\lambda }
	=
	\frac{ q(x)k^v }{ 2r } \frac{d}{dx}.
\ee
With this, equation \eqref{tangent} is rewritten as
\be
	\frac{d k^v}{dx}
	+
	\frac{2\mu(d-3)x^{d-3}}{q(x)} k^v = 0.
\ee
A formal solution of this equation is
\be
	\frac{ k^v }{ k^v_0 }
	=
	\exp\left[
		\int_0^x
		K(x)dx
	\right],
\;\;\;
	K(x):= - \frac{ 2\mu (d-3)x^{d-3} }{ q(x) },
\label{k/k0}
\ee
where $k^v_0 = k^v|_{x=0}$ is an integration constant.

First, let us consider the generic case in which the algebraic equation $q(x)=0$ has the non-degenerate roots at $x=x_-$ and $x=x_+$. Subtracting a pole of the integrand in equation \eqref{k/k0}, we have
\begin{align}
\begin{split}
	\frac{k^v}{k^v_0}
	&=
	\exp\left[
		-\int_0^x
		\frac{ 2\mu(d-3)x_-^{d-3} }{ q^\prime(x_-) ( x-x_- )}
		dx
	\right]
	\exp\left[
		\int_0^x
		\left(
			K(x) +
			\frac{ 2\mu(d-3)x_-^{d-3} }{ q^\prime(x_-) ( x-x_- )}
		\right)
		dx
	\right]
\\
	&=
	\left( \frac{x_- - x}{x_-} \right)^\beta
	\exp \left[ \int_0^x K_\ast (x) dx \right],
\end{split}
\label{kv/k0_generic}
\end{align}
where
\be
	K_\ast(x)
	:=
	K (x) + \frac{ 2\mu(d-3)x_-^{d-3} }{ q^\prime(x_-) ( x-x_- )},
\;\;\;
	\beta
	:=
	- \frac{2\mu(d-3)x_-^{d-3}}{ q^\prime(x_-) }
	=
	\frac{ x_- - 2 }{ x_c - x_- }.
\ee
$K_\ast(x)$ is finite at $x=x_-$ and the last integral in equation \eqref{kv/k0_generic} takes a finite value in the limit $x \to x_- - 0$. Note that $\beta = \alpha -1 $ ($>0$) holds, where $\alpha$ is the power of the map in equation \eqref{G}. This relation $\beta = \alpha -1 $ implies that the divergence of the power and energy (especially, in the minimally coupling case) stems from the divergence of the redshift at the Cauchy horizon, which can be seen from equation \eqref{kv/k0_generic}.

Next, we consider the critical case in which $q(x)=0$ has the degenerate root at $ x = x_c $. In a similar way to that of the generic case, we subtract a pole of the integrand, which is second order in this case,
\begin{align}
\begin{split}
	\frac{k^v}{k^v_0}
	&=
	\exp \left[
		- \int_0^x
		\frac{ 2\mu (d-3)x_c^{d-3} }{ (1/2)q^{\prime\prime}(x_c)  (x-x_c)^2 } dx
	\right]
	\exp \left[ \int_0^x
		\left(
			K(x) + \frac{ 2\mu (d-3)x_c^{d-3} }{ (1/2)q^{\prime\prime}(x_c)  (x-x_c)^2 }
		\right) dx
	\right]
\\
	&=
	\exp\left[
		- \frac{4}{(d-3)(x_c-x)}
	\right]
	\exp
	\left[
		\frac{2}{d-2} + \int_0^x \hat{K}_\ast(x) dx
	\right],
\end{split}
\end{align}
where $\hat{K}_\ast(x)$ is a function regular at the Cauchy horizon $x=x_c$,
\be
	\bar{K}_\ast(x)
	:=
	K(x) + \frac{ 2\mu (d-3)x_c^{d-3} }{ (1/2)q^{\prime\prime}(x_c)  (x-x_c)^2 }.
\ee

%----------------------------------------------------------------------%
%----------------------------------------------------------------------%

%----------------------------------------------------------------------%
%----------------------------------------------------------------------%

\end{document}